\documentclass[epsfig,graphics,twocolumn,floatfix,a4paper,showpacs]{revtex4}
\usepackage{amsmath,amsfonts,amssymb,graphicx,color,times,bbm,psfrag,dsfont}
\usepackage[english]{babel}
\usepackage[latin1]{inputenc}
\usepackage[T1]{fontenc}

\newcommand{\dd}{\ensuremath{\textrm{d}}}
\def\one{\ensuremath{\hbox{$\mathrm I$\kern-.6em$\mathrm 1$}}}

\begin{document} 

\title{Area law and vacuum reordering in
harmonic networks}

\author{A. Riera and J. I. Latorre}

\affiliation{
Dept. d'Estructura i Constituents de la Mat\`eria,
Univ. Barcelona, 08028, Barcelona, Spain.
}
\date\today

\begin{abstract}

We review a number of ideas related to area law scaling
of the geometric entropy 
from the point of view of condensed matter, quantum field theory and
quantum information. 
An explicit computation in arbitrary dimensions of 
the geometric entropy of the ground state of a
discretized scalar free field theory shows the expected
 area law result. In this case, area law scaling is a manifestation of
a deeper reordering of the vacuum 
produced by majorization relations.
Furthermore, the explicit control on all the eigenvalues of the
reduced density matrix allows for a verification
of entropy loss along the renormalization group 
trajectory driven by the mass term. A further result of
our computation shows that single-copy entanglement also obeys
area law scaling, majorization relations and decreases along renormalization
group flows.  
\end{abstract}
\maketitle

\section{Introduction}

The amount of entanglement present in a quantum state is
of fundamental relevance to determine how
hard it is to simulate it by classical means. 
It is generally argued that a highly entangled quantum 
state carries a huge superposition
of product states that cannot be handled on a classical computer.
Yet, this statement must be made precise, since a small amount
of entanglement can indeed be simulated efficiently. 
The relevant precise question is, thus, how much entanglement can be
efficiently simulated classically.

This abstract question should at least be clarified 
when considering relevant physical systems. Can the
amount of entanglement present in a  two-dimensional 
lattice of harmonic oscillators
be efficiently represented in a classical computer?
Although the answer to this question is  not yet settled, qualitative 
progress has been recently achieved. One of the  ingredients
essential to this discussion is the area law for
the geometric entropy and the representation of
quantum states by projected entangled pairs. 

A important related problem is to understand how
entanglement varies along renormalization group (RG)
trajectories. We shall bring growing evidence for
the idea that RG
flows entail a loss of entanglement. This 
entanglement loss will be shown 
compatible with area law scaling of the entropy. 

We organize the contents of this paper by first reviewing
a number of previous results on area law scaling of 
ground state entropy
in different systems using the language of
condensed matter, quantum field theory and
quantum information theory. We shall then present a
computation of entanglement entropy on a discretized
bosonic free field theory in arbitrary dimensions. This
gives us control on the eigenvalues of the 
reduced density matrix on a subsystem which,
in turn, allows for a discussion of majorization
relations obeyed by the reduced density matrix
of the system. We extend this discussion
to the single-copy entanglement measure. RG loss of
entanglement is also verified in detail
for arbitrary dimension networks of harmonic
oscillators.

\section{A brief review of the area law}
\subsection{Measures of entanglement for many-body systems}
An arbitrary quantum  state of {\sl e. g. }
$N$ spins is, in general, highly entangled.
To quantify such a statement we can use 
various figures of merit. For instance, 
concurrence \cite{Wo98} is easy to compute
and detects some pairwise entanglement though it cannot 
scan correlations throughout the system. 
An appropriate and widely used candidate to quantify entanglement is
the von Neumann entropy of the reduced density matrix
of the state under analysis, when only a subset of degrees of freedom
is retained. To be precise, let us consider a quantum state
made out of $N$ qubits $ \vert \psi\rangle \in (C^2)^{\otimes N}$
and its density matrix of a $\rho\equiv \vert \psi\rangle\langle\psi\vert$.
Next, we consider the reduced density matrix of a subset of qubits
denoted by $A$,
$ \rho_A\equiv {\rm Tr}_{\bar A}\rho$ 
where all qubits but those belonging to
the set $A$ are traced out. The von Neumann entropy of
the reduced density matrix is then defined as
\begin{equation}
 \label{reducedensitymatrix}
 S(\rho_A)=-{\rm Tr} \rho_A \log\rho_A\ .
\end{equation}
The entropy is often referred to as {\sl entanglement entropy}. 

In general a set of particles will be distributed
randomly over space. Entanglement entropy can be
computed for all sorts of partitions of the system,
yielding information about the quantum correlations
among the chosen subparts. A particular and extremely relevant
class of physical systems are those made of local quantum degrees
of freedom which are arranged in chains or, more generally,
in networks. For such systems it is natural
to analyze their entanglement by studying
geometrical partitions, that is, computing the
entanglement entropy between a set of contiguous qubits
versus the rest of the system. We shall referred to this particular
case of entanglement entropy \cite{Sr93} 
as {\sl geometric entropy} \cite{CW94} (also
called {\sl fine grained entropy} in \cite{FPST94}).

 The appearance of scaling of the geometric entropy
with the size of the sub-system under consideration
has been shown to be related to quantum phase transitions
in one-dimensional systems, further
reflecting the universality class corresponding
to the specific phase transition under consideration
\cite{CW94,VLRK} (see also \cite{AOPFP04}). 
Broadly speaking, a large
entropy is related to the presence of long distance correlations,
whereas a small entropy is expected in the presence of
a finite correlation length.
The precise scaling of geometric entropy does eventually 
determine the limits for today's efficient simulation of a 
physical quantum  system
on a  classical computer. 

There are 
many other ways to quantify entanglement. The von Neumann
entropy we have chosen as our central figure of merit
has an asymptotic operational meaning. Given infinitely many 
copies of a bipartite quantum state, it quantifies
how many EPR pairs can be obtained
using local operations and classical communication.
 A different measure of entanglement 
can be associated with the analysis of entanglement on a single-copy 
of the quantum system. This single-copy entanglement 
\cite{EC05,PZ06,OLEC05,KMSW06}
can be defined as
\begin{equation}
  \label{e1L}
  E_1(\rho_A)=-\log \rho_A^{(1)}
\end{equation}
where $\rho_A^{(1)}$ is the maximum eigenvalue of $\rho_A$.
This quantity provides the amount of maximal entanglement 
that can be extracted from a single copy of a state
by means of local operations and classical communication.
As we shall see later on, the von Neumann entropy and the 
single-copy entanglement appear to be deeply related in any 
number of dimensions.

\subsection{Volume {\sl vs.} area law}
Random states are known to carry large entanglement. To be
precise, let us consider a random
infinite system o qubits. On average, 
the density matrix for a random
subset of $N$ qubits carries maximum
von Neumann entropy, 
\begin{equation}
\label{bulke}
 S(\rho_N)\sim N \ .
\end{equation}
This result \cite{PageZ} shows that the entropy of random states
grows as the number of particles included in the subset.
This is referred to as a {\sl volume law} scaling. An arbitrary
state uses the maximum possible superposition of
the basis elements with no symmetry whatsoever
among their coefficients. Its efficient representation
by classical means
appears certainly difficult.

Physical theories create entanglement through
interactions, which are typically local. Thus,
{\sl e.g.} the ground state of a sensible physical
Hamiltonian is not a random state. It is
natural to expect a low amount of entropy
since local interactions will entangle the 
non-contiguous degrees of freedom in a somewhat sequential
way. We may encounter local intense 
entanglement that dilutes at long distance.
This is precisely the structure of standard
quantum theories, with correlations that decay
with a power law at phase transitions and with 
an exponential law
away from them. It is then
reasonable to ask what is the limit of
efficient simulability in terms of the 
entanglement present in a given state.

In many physical theories, local degrees of freedom 
are arranged in a specific geometrical way as mentioned previously. 
We may have
quantum systems defined on spin chains, 
networks or, in general, $D$-dimensional
lattices. Those systems may have a 
continuum limit described by a quantum field theory
or, alternatively, may be devised as quantum
simulators, a preview of quantum computers.
We may then discuss the amount of geometrical entanglement
present on the system from three complementary points of view:
condensed matter, quantum field theory and quantum information. 

As we shall see, the basic ingredient of locality of
interactions suggests that entropy
for a geometrical region should be dominated by
the entanglement present on
the surface separating it from the
rest of the system.
To be precise, 
consider an infinite $D$-dimensional lattice
where we  assign  part $A$ to an inner hypercube of
size $L$, $N=L^D$, and part $B$ to the outside. Locality seems 
to suggest
\begin{equation}
\label{arealaw}
S(\rho_L)\sim L^{D-1}\sim N^{\frac{D-1}{D}} \ .
\end{equation}
This behavior is commonly referred to as  {\sl area law} scaling
for the geometric entropy. 
Let us note that one-dimensional quantum systems 
correspond to  a well understood limiting case 
for the above formula, where the 
power law turns out to be substituted with a logarithmic
scaling at phase transitions, that is 
\begin{equation}
\label{arealawonedimension}
S(\rho_L)\sim \log L  ,
\end{equation}
and saturates away from them
\begin{equation}
\label{arealawonedimension}
S(\rho_L)< {\rm constant}\quad,\forall L \quad ,
\end{equation}
as shown in Ref. \cite{VLRK,JK04,CC04}
These results are deeply connected
to  conformal symmetry and control
the classical simulability of the system.

Recent evidence hints at
 a log violation of the area law in some two-dimensional
 systems made with anticommuting variables \cite{Wo05,GK05,LLYRH06,BCS06}.
To be precise, some of these models display
an entropy scaling law of the type
\begin{equation}
\label{logarea}
S(\rho_L)\sim L^{D-1} \log L .
\end{equation}
It is unclear whether such systems support
a limiting quantum field theory description
in the continuum limit.

It is important to
make a general remark concerning
the different approaches
to the computation of entanglement in quantum
systems. Let us note that discretized quantum systems
allow for uncontroversial computations of the
entropy. This is not the case of
quantum field theories, where regularization and
renormalization are needed since
the number of degrees of freedom
is formally unbounded. In such a framework, 
the adimensional
entropy requires the appearance of some
short-distance regulator $\epsilon$
\begin{equation}
\label{arealawqft}
S(\rho_L)\sim \left(\frac{L}{\epsilon}\right)^{D-1}
\end{equation}
which entails the
necessary discussion of its renormalization and its
observability. Let us just mention here that
the coefficient of the area law is universal
for $D=1$ systems whereas remains scheme-dependent
in higher dimensions.

The problem turns extremely subtle in the case of gravity,
where the geometry of space-time is dynamical and
the way to compute for a black hole the Bekenstein
area law pre-factor from first principles
is far from clear\cite{Fursaev,RT06,E06}. Recent progress on the side
of AdS/CFT correspondence seems to link 
entanglement entropy in a quantum field theory
living on the boundary to the black-hole
entropy of the bulk \cite{RT06,E06}.

\subsection{Locality and PEPS}
The basic heuristic argument for an area law scaling of entropy
for the ground state of physical systems is rooted in the locality of
the interactions. Steps to make this argument
quantitative have been made in Refs. \cite{VC04,VPC04,PEDC05,VC05,H05,VWPC06}.

A local Hamiltonian tends to 
entangle nearest neighbors. Long-distance entanglement
emerges as a coherent combination of local interactions. The
correctness of this argument would imply that
the reduced entropy of a geometric bipartition of
a system will get its main contribution from
the entanglement between degrees of freedom
at opposite sites of the boundary that separates
the regions. This, in turn, implies an area-law
scaling. Let us note that such a naive argument
works in any dimension and does not depend on
the correlation length present in the system.
Area law would emerge from locality,
whatever the mass-gap is. We shall
discuss the limitations of this argument shortly.

This argument needs a clear formulation and
verification. Although we lack definite answers
about the necessary and sufficient conditions
a Hamiltonian must obey to produce a ground
state with area law entropy, some  progress has been
achieved using one-dimensional
Matrix Product States (MPS) and their generalization
to higher dimensions,
Projected Entangled Pair States (PEPS). We first consider
a one-dimensional system with open
boundary conditions described by a MPS
\begin{equation}
\label{PEPs}
\vert \psi\rangle =
\sum \left( A^{i_1}_{\alpha_1}A^{i_2}_{\alpha_1\alpha_2}
\dots A^{i_n}_{\alpha_{n-1}}\right)
\vert i_1 \dots i_n\rangle \ ,
\end{equation}
where the sum extends to $i_1,\dots,i_n=1,\dots,d$, 
which are physical indices attached to 
local Hilbert spaces, and $\alpha_1,\dots,\alpha_{n-1}=1,\dots \chi$, which
are ancillae indices. The tensors $A^i_{\alpha\beta}$ can be viewed
as projectors from the ancillae indices to a physical one. This representation
provides the basis for the density matrix renormalization group
technique.

The generalization of the MPS construction to 
higher dimensional networks carries the
name of PEPS. In a $D$-dimensional network, 
where ancillae degrees of freedom
are linked to their nearest neighbors, the role of
the MPS projector is taken by a tensor of the form
\begin{equation}
\label{PEPs}
A^{a}_{\alpha{\gamma \atop\delta}\beta} \ ,
\end{equation}
where the physical indices span
a $D$-dimensional lattice  and ancillae run from 1 to $\chi$.
Again, the role of each tensor $A$ is to project
maximally entangled pairs connecting  
local neighbors onto a physical local space. 
Entanglement is thus carried by the links connecting ancillae.
Each entangled pair, that is, each sum over one ancilla index hides
a connecting bond of the type $\sum_{\alpha=1}^\chi
\frac{1}{\sqrt \chi}\vert\alpha\alpha\rangle$.
If one of the two ancillae in the bond is traced
out, the entropy for the remaining ancilla is
$S=\log \chi$.

We are now in a position to present the
argument in Ref.  \cite{VC04}
showing  that finite $\chi$ PEPS
entail area law scaling for the entropy.
Let us assume that the ground state of a quantum system
is described by a PEPS with finite $\chi$. It follows
that the entropy of a subpart of the system is bounded
by the number of bonds which are cut by the separating surface
times the entropy per broken bond. This amounts to
an area law
\begin{equation}
\label{peps bond}
S(\rho_A)\leq (\#{\rm cut\ bonds}) \log \chi \sim {\rm Area} \log \chi  .
\end{equation}
A violation of the area law within the PEPS representation requires
infinite-dimensional ancillae. 

We should again distinguish the one-dimensional case,
 where the ground state of infinite critical systems 
are known to carry logarithmic entropy \cite{VLRK}, 
\begin{equation}
\label{onedimension}
S(\rho_L)\sim \frac{c}{3}\log \frac{L}{a} \ ,
\end{equation}
where $a$ is the lattice spacing
and $c$  the central charge that characterizes the 
universality class of the phase transition.
Yet, the
boundary of a one-dimensional block is made by two single points.
Such a state with logarithmic entropy 
cannot be represented using finite dimensional
MPS and we must resort to arbitrarily large $\chi$. 
This limitation is at the heart of the problems
that the DMRG technique encounters when applied to 
quantum phase transitions. On the other hand,  
the entropy is bounded away from critical points and MPS provide an
efficient way to represent the system. MPS states
with finite $\chi$ are often referred to as finitely correlated
states.

Coming back to higher dimensions, 
it is then a major issue to establish whether 
finite $\chi$ PEPS can describe faithfully the ground
state of physical systems. The fact that PEPS
with finite $\chi$ can
incorporate an area law is appealing. Recently, 
 a particular class of finite 
PEPS has been constructed that display 
polynomial decay laws, that is long range correlation
\cite{VWPC06}.
These PEPS are also shown to describe ground states
of frustration-free Hamiltonians and such
states can approximate exponentially well any 
finitely correlated state.
It is still unclear whether the ground states of
standard quantum systems fall into this description or, alternatively,
they need infinite $\chi$. This may set apart  what
is efficiently simulable from what is not.
 
\subsection{Renormalization group transformations on MPS and PEPS 
and the support for an area law}

We have argued that one-dimensional finite $\chi$ 
MPS can support a maximum
amount of entropy independent of the size of the system
and that, in contradistinction, finite D-dimensional PEPS can
accommodate an area law. Let us give an independent
quantitative argument for this statement.

Consider a renormalization group transformation  
of a MPS state with constant $A$ defined by the coarse graining of two
sites \cite{VCLRW04}
\begin{equation}
\label{RG1}
A^i_{\alpha\beta}A^{j}_{\beta\gamma}\equiv \tilde A^{ij}_{\alpha\gamma}=
\sum_{l=1}^{\min (d2,\chi2)}
\lambda_l U^{(ij)}_l V^l_{\alpha\gamma}
\end{equation}
where we have decomposed the product of two
adjacent matrices using a singular value decomposition.
We can understand the unitary matrix $U$ as a 
change of basis on the new coarsed degree of freedom
and construct a new MPS with $A'^l_{\alpha\gamma}=\lambda_l 
V^l_{\alpha\gamma}$.
Therefore, the ancillae indices close under such operation
whereas the physical index grows.
 Upon iteration
of this operation, the range of the physical index will reach
a maximum value $\chi2$ and will get locked to that value.
This is the magic of one dimension. The long-distance
properties of the system are completely described by a 
single square effective matrix! Entropy is then
bounded.

The analogous argument in two-dimensional systems
follows a slightly different path. The coarse graining step
reads
\begin{eqnarray}
\label{RG2}
&{A^{a}_{\alpha{\gamma{\phantom '} \atop\nu{\phantom '}}\mu} 
\atop 
A^{c}_{\alpha'{\nu{\phantom '} \atop\delta{\phantom '}}\mu'}} 
{A^{b}_{\mu{\gamma' \atop\nu'}\beta} 
\atop 
A^{d}_{\mu'{\nu' \atop\delta'}\beta'}} =\tilde A^{{ab\atop
    cd}}_{\alpha\alpha'{\gamma\gamma'\atop\delta\delta'}\beta\beta'}
\\ &
=\sum_{l=1}^{\min (d4,(\chi4)2)}\lambda_l U^{{ab\atop cd}}_l
V^l_{\alpha\alpha'{\gamma\gamma'\atop\delta\delta'}\beta\beta'}
\end{eqnarray}
As before, we can absorb the global $U$ as a change of the local
coarse-grained basis and assign a new PEPS to $\lambda V$.
Note the different growth of indices. 
On the one hand, physical indices merge in groups of four
and would naively need a volume law increase, $d4$. 
On the other hand, the ancillae rank  increase from $\chi4$ to 
$(\chi4)2$, that is, it follows an area law.
Given that the singular value decomposition will
be locked by the smallest dimension of the two above,
the area law will define the rank of the tensor
that contains the effective
long distance description of the model. 
The argument generalizes to $D$ dimensions
where the PEPS $A^i_{\alpha_1,\dots,\alpha_{2D}}$
with a physical index $i=1,\dots,d$ and ancillae
indices $\alpha_1,\dots,\alpha_{2D}=1,\dots, \chi$.
A renormalization group transformation of this PEPS 
makes the new collective physical index to run
$i'=1,\dots, d^{2^D}$, that is, with a volume law,  and
the new collective ancillae $\alpha'_1,\dots,\alpha'_{2D}=
1,\dots, \chi^{2^{D-1}}$, that is, as an area law.
The singular value decomposition makes all the long-distance
properties of the state to be contained in an effective
PEPS with a number of degrees of freedom that grows
with just an area law. The rank of the
effective PEPS is $\log \chi_{eff}=2^{D-1}\log\chi$.
From this simple argument, 
it follows that PEPS can support an area law scaling for
the geometrical entropy.

\subsection{Some explicit examples of area law}
There is an extensive literature on computations
of the entropy for particular cases that cannot
be faithfully summarized here.

One-dimensional  spin systems ({\sl e.g.}
quantum Ising model, XX model and Heisenberg model) 
obey a logarithmic scaling at the critical point 
\cite{JK04,CC04,AEPW02,VLRK,IJK04,Pe04}.
Away from the quantum phase transition point, the
entropy gets saturated. This explicit computation
falls into the universal scaling predicted by
conformal invariance. This result
has been further verified and extended to 
many other quantum systems in one dimension.
I

The literature on computations of entanglement
entropy in higher dimensional systems is far
less extensive due to the difficulty to produce
explicit results.
The first analysis of the entanglement entropy
in two- and three-dimensional systems were
done in discretized approaches to quantum field
theory \cite{Sr93,FPST94,BKLS86}. Further
analysis showed that the entanglement entropy is 
related to the trace anomaly in curved space times
giving an explicit relation between the actual
results for free fermions and free bosons \cite{KS94}. 

Rigorous computations in 
discretized harmonic networks proved
no departure from the area law \cite{PEDC05, CrEi}. 
Further analysis of entanglement entropy on higher dimensional
networks has been done in Refs. \cite{BR04, HIZ05,DHHLB05}.

\subsection{Exceptions to the area law}
We have argued that the area law is deeply connected to 
locality of interactions. It is, therefore, reasonable
to expect violations of such scaling in models with
non-local interactions. This is the generic case
of a quantum computation of an NP-complete problem.
It has been numerically verified that this is the
case when an adiabatic quantum computation is
applied to the NP-complete Exact Cover problem, a variant of
the 3-SAT problem. Along the computation, the
ground state becomes maximally entangled, that is
its entropy scales as the volume of the system 
\cite{OL04,BOLPR05}.
A physical quantum computer will definitely need to
face the challenge of maintaining those huge 
fine-tuned superpositions of states. 

\begin{table}[t]
\begin{center}
\begin{tabular}{cc}
\hline
Spin chains away from criticality& $S\sim {\rm constant}$\\
Critical spin chains& $S\sim \log N$\\
$D$-dimensional harmonic networks& $S\sim N^{\frac{D-1}{D}}$\\
NP-complete problems & $S\sim N$
\\
\hline
\end{tabular}
\end{center}
\end{table}

Locality of interactions is not the only ingredient
that controls entropy. Entropy is related to the
eigenvalues of the Schmidt decomposition of a system
in two parts. If the subsystems retain a lot of
symmetry, the sub Hilbert spaces organize themselves
in representations of the symmetry group. This
entails a reduction of the Schmidt number of the
above decomposition, that is, a lower entropy. 
Such a counter mechanism to reduce the
entropy in highly connected systems has been explicitely checked
in the case of the Lipkin-Meshkov model which
is defined by a spin system fully and symmetrically connected.
Although it is tempting to argue that the system
is infinite-dimensional (the geometry of the Hamiltonian
corresponds to a simplex of $N\to \infty$ vertices), 
the entropy  scales only logarithmically,
which is the actual bound for symmetric spaces
\cite{LORV04}. This logarithmic scaling of the entropy
follows the one-dimensional log law, which
might just be an accident.

It should not come as a surprise that 
slightly entangled states that do not
correspond to an eigenstate of a given Hamiltonian
dynamically evolve
to highly entangled states under its action. This has been
analyzed in Refs. \cite{CC05, CMCF06}
even for simple Hamiltonians like the quantum Ising chain.
No area law is expected for 
slightly entangled random states
when they are evolved with local Hamiltonians.

As mentioned previously,  a
case of non-trivial violation of the area law
was first considered in \cite{Wo05}
and then analyzed in \cite{GK05,LLYRH06}. 
Some two-dimensional systems with anticommuting variables
were found to display a log correction to
the area law, that is, $S\sim L \log L$. 
On the other hand, some previous computations for
free Dirac fermions seem to produce no area-law
violation \cite{LW95,Ka95,CFH05} in any number of
dimensions. This issue
deserves further investigation.
Finally let us mention that in the computation 
of quantum corrections to the entropy of a black hole, 
logarithmic corrections have also been obtained \cite{S95}.

\subsection{Physical and computational meaning of an area law}
We can attach physical meaning to an area law scaling of
entropy in different but related ways. We may argue that
entropy is a measure of surprise due to quantum correlations and
that a state that obeys an area law carries less correlations than
a random state. As the size of the  inner block increases,
we only get a reduced amount of surprise, compare
to the maximum possible,  when
discovering that our block was correlated to the exterior.
It is then
arguable that the theory that has produced such a state may
accept a simpler description. In some sense, this argument
is implicit in the holographic description of some
quantum systems.

From a computational point of view, low entropy means small
quantum correlations, that is, small entanglement. It is known
that states that are only  slightly entangled can
be efficiently simulated by classical means
\cite{Vi03}.
A fundamental
question is thus formulated: what entropy growth law 
can be efficiently simulated by a classical computer? 

So far, this question can only be answered partially.
In one dimension, $D=1$, quantum critical phenomena
show a logarithmic scaling  which cannot be
reproduced using finite MPS techniques. Formally,
the simulation remain efficient in the sense that
to reproduce critical behavior we need $\chi$
to be polynomial in $L$. This, though, produces
an obvious practical computational slowing down
and limitation. A new promising idea  to represent a
quantum state with a  different and non-local tensor structure
 has been proposed 
in Ref. \cite{Vi05} with the name
of multi-scale entanglement renormalization ansatz
(MERA). The basic idea is to substitute a linear
MPS representation with a RG-inspired construction
that also identifies the key use of
disentangling operations for blocks before proceeding
to a coarsed description. 

The question in two dimensions has been addressed
in \cite{VC04} in a sequential
way. A PEPS is taken as lines of spins that are collected into
effective degrees of freedom which are further
treated in a MPS manner. 

\section{Area law in $D$ dimensions}
\subsection{The Hamiltonian of a scalar field in $D$ dimensions}
Let us consider the theory of 
a set of harmonic oscillators in $D$ dimensions which
is expected to verify area law scaling of the entropy.
A number of non-trivial issues can be discussed in 
this explicit example. First, we shall analyze
the regularized version of a scalar free field theory
in order to get its reduced density matrix when
an inner geometrical ball is integrated out.
Its eigenvalues can, then, be used to compute the
geometrical entropy that will scale as dictated
by the area law. Second, we can compare the 
behavior of the entropy to the one of the 
single-copy entanglement. Third, we can 
analyze whether area law scaling is backed
by a deeper sense of order, namely majorization
theory.

Our computation will generalize the one presented 
in Ref. \cite{Sr93} to $D$ dimensions. Let us consider
the Klein-Gordon Hamiltonian
\begin{equation}
  H=\frac{1}{2}\int \dd^Dx \left(\pi^{2}(\vec{x}) 
    +\left|{\nabla \phi(\vec{x})}\right|^2 
    +\mu^{2}\left|\phi(\vec{x})\right|^2 \right)\ ,
 \label{hamiltoniaD0}
\end{equation}
where $\pi(x)$ is the canonical momentum associated to the
scalar field $\phi(x)$ of mass $\mu$.
The $D$-dimensional Laplacian reads
\begin{equation}
\Delta \phi=\frac{1}{r^{D-1}}\frac{\partial}{\partial{r}}
  \left(r^{D-1}\frac{\partial{\phi}}{\partial{r}}\right)+
  \frac{1}{r^{2}}{\hat L}^{2}\phi\ ,
\end{equation}
where $r=|\vec{x}|$ and ${\hat L}^2$ is the total angular momentum operator
in $D$ dimensions. It is convenient 
to introduce the real spherical harmonic functions $Z_{l\{m\}}$, which
are eigenfunctions of ${\hat L}^{2}$ with eigenvalues $l(l+D-2)$. The set of 
numbers $\{m\}$  stand for other Casimir and component labels 
in the group $SO(D)$. We now  project 
the angular part of the scalar fields $\pi$ and $\phi$,
\begin{subequations}
 \begin{align}
 \pi_{l\{m\}}(r)&=r^\frac{D-1}{2}
 \int \dd^Dx Z_{l\{m\}}(\theta_1,\cdots,\theta_{D-2},\varphi)\pi(\vec{x})  \\
 \phi_{l\{m\}}(r)&=r^\frac{D-1}{2}
 \int \dd^Dx Z_{l\{m\}}(\theta_1,\cdots,\theta_{D-2},\varphi)\phi(\vec{x}) \ ,
\end{align} 
\label{eigenbasisD}
\end{subequations}
where $r,\theta_1,\cdots,\theta_{D-2}$ and $\varphi$ define
 the spherical coordinates in $D$ dimensions. 
The Hamiltonian now reads
 \begin{equation}
  H=\sum_{l\{m\}} H_{l\{m\}}
  \label{hamiltonia1}
  \end{equation}
where,
\begin{align}
  H_{l\{m\}}&=\frac{1}{2}\int \dd r \left(\pi_{l\{m\}}^2(r)+
  r^{D-1}\left(\frac{\partial}{\partial{r}}
  \left(\frac{\phi_{l\{m\}}(r)}{r^{\frac{D-1}{2}}}\right)\right)^{2} \right. \nonumber \\
  & \left.+\left(\frac{l(l+D-2)}{r^{2}}+\mu^{2} \right)\phi_{l\{m\}}^2(r) \right)
  \ .
\label{Hlm_definiton}
\end{align}
An ultraviolet regularization of the 
radial coordinate in the above Hamiltonian will transform the scalar
field theory into a chain of coupled harmonic oscillators. 
This is achieved by discretizing  
the continuous radial coordinate $r$ into a lattice of $N$ discrete 
points spaced by a distance $a$,
\begin{align}
  H_{l\{m\}}&=\frac{1}{2a}\sum_{j=1}^N \left( \pi_{l\{m\},j}^2 \right. \nonumber \\
   & \left. +(j+\frac{1}{2})^{D-1}
   \left(\frac{\phi_{l\{m\},j+1}}{(j+1)^{\frac{D-1}{2}}}-  
    \frac{\phi_{l\{m\},j}}{j^{\frac{D-1}{2}}}\right)^{2}  \right. \nonumber \\
    & \left. +\left(\frac{l(l+D-2)}{j^{2}}+\mu^{2} \right)\phi_{l\{m\},j}^2(x)\right) \ .
  \label{hamiltonia2}
\end{align}
 The size of the system is 
$L=(N+1)a$, where $a$ and $L$ act as an ultraviolet 
and infrared cutoff respectively.
We can compare this expression with the Hamiltonian 
of an open chain of  $N$ coupled harmonic oscillators,
\begin{equation}
H=\frac{1}{2}\sum_{i=1}^N p_i^{2} +\frac{1}{2}\sum_{i,j=1}^Nx_iK_{ij}x_j
\end{equation}
and identify $K_{ij}$ as 
\begin{align}
 \label{Kdefinition}
 K_{ij}&= \left(\frac{l(l+D-2)}{j^{2}}+\mu^{2}\right)\delta_{ij} \nonumber \\
 &+\left(1-\frac{1}{2j} \right)^{D-1}\theta\left(j-\frac{3}{2}\right)\delta_{ij}  \nonumber \\
 &+\left(1+\frac{1}{2j} \right)^{D-1}\theta\left(N-\frac{1}{2}-j\right)\delta_{ij} \nonumber \\
 &+\left(\frac{j+\frac{1}{2}}{\sqrt{j(j+1)}}\right)^{D-1}\delta_{i,j+1}+
 \left(\frac{i+\frac{1}{2}}{\sqrt{i(i+1)}}\right)^{D-1}\delta_{i+1,j} \ ,
\end{align} 
where $\theta$ is the step function.
\subsection{Geometric entropy and single-copy entanglement}
\label{chain of oscillators}
We now proceed to trace out an inner geometric  ball around the origin
to obtain the reduced density matrix of the ground state of the 
system on the exterior
of that ball. Following similar steps as in \cite{Sr93}
we define $\Omega$ as the square root of $K$, that is  $K=\Omega2$. 
The gaussian ground state of the system can be expressed as,
\begin{equation}
\psi_0(x_1,...,x_N)=\pi^{-N/4}(\det
\Omega)^{1/4}e^{-\frac{x^T\cdot\Omega\cdot x}{2}}\ ,
\end{equation}
where $x\equiv(x_1,...,x_N)$. We construct the density matrix 
 $\rho_{out}$ by tracing over the inner $n$ oscillators,
\begin{equation}
\rho_{out}(x,x')\sim e^{-\frac{1}{2}(x^T\cdot\gamma\cdot x + 
x'^T\cdot\gamma\cdot x')+x^T\cdot\beta\cdot x'}\ ,
\end{equation}
where  $\beta$ and $\gamma$ are defined by
\begin{subequations}
\begin{align}
\beta &\equiv \frac{1}{2} B^TA^{-1}B \\
\gamma &\equiv C-\beta
\end{align}
\end{subequations}
and $A=\Omega(1{\div}n , 1{\div}n)$, $B=\Omega(1{\div}n,n+1{\div}N)$ and 
$C=\Omega(n+1{\div}N,n+1{\div}N)$ are sub-matrices of 
$\Omega$.

We proceed with the diagonalization of this structure
rotating and rescaling the variables $x=V^T\gamma_D^{-1/2}y$ 
where $\gamma=V^T\gamma_DV$  and $\gamma_D$ is diagonal. Using this 
transformation, $\gamma$ becomes identity, 
$\beta \to \beta'=\gamma_D^{-1/2}V\beta V^T\gamma_D^{-1/2}$ 
and the density matrix reads
\begin{equation}
  \rho_{out}(y,y')\sim e^{-\frac{1}{2}(y^2+y'^2)+y^T\cdot\beta'\cdot y'}\ .
\end{equation}
If we do the appropriate change of coordinates  $y=W\cdot z$ 
(where $W$ is an orthogonal matrix) 
such that $W^T\cdot \beta'\cdot W$ becomes 
diagonal with eigenvalues $\beta_i'$, we get $\rho_{out}$ as a tensor 
product of the two coupled harmonic oscillators density matrices, 
\begin{equation}
  \rho_{out}(z,z')\sim \prod_{i=1}^{N-n} e^{-\frac{1}{2}(z_i^2+z_i'^2)+\beta_i'
  z_i z_i'}\ .
 \label{tracedDM}
\end{equation}

We can now compute the entropy 
associated to the reduced density matrix $\rho_{out}$.
This entropy can be expressed as a sum over 
contributions coming from each term in the 
reduced density matrix tensor
product structure,
\begin{equation}
  S_{l\{m\}}=\sum_{i=1}^{N-n} S_{l\{m\},i}(\xi_i)\ ,
\label{partial_entropy1}
\end{equation}
where
\begin{equation}
  S_{l\{m\},i}(\xi_i)=-\log(1-\xi_{l\{m\},i})-
  \frac{\xi_{l\{m\},i}}{1-\xi_{l\{m\},i}}\log\xi_{l\{m\},i}
  \label{partial_entropy0}
\end{equation}
is the entropy associated to each sub-density matrix in the 
product shown 
in Eq.(\ref{tracedDM}) and $\xi_{l\{m\},i}$ 
is the parameter that generates the eigenvalues  
of these densities matrices. Note that each eigenvalue $\xi=\xi_{l\{m\},i}$
entails a set of probabilities of the form 
\begin{equation}
\label{spectra}
p_{n}=(1-\xi)\xi^{n} \ \ \ \ n=0, 1, 2, 3,\ldots \ \,
\end{equation}
defined by $\xi_i=\beta_i'/(1+(1-\beta_i'^2)^{1/2})$ for
each $l\{m\}$ set. 

To compute the total entropy, we have to sum 
over all possible values of $\{m\}$ and $l$.
\begin{equation}
  S=\sum_{l\{m\}} S_{l\{m\}}\ .
  \label{entropy1}
  \end{equation}
We realize from Eq.(\ref{Hlm_definiton}) that $H_{l\{m\}}$ only 
depends on $l$, so the entropy associated to its ground state will 
also be $\{m\}$ independent, and therefore
\begin{equation}
  S=\sum_{l=0}^{\infty} \nu(l,D)\, S_{l} \ ,
  \label{entropy2}
  \end{equation}
being $\nu(l,D)$ the degeneracy of the total angular momentum 
operator $\hat L^2$ for a fixed $l$. 
In three dimensions, for example, $\{m\}=m$ can go from $-l$ to $l$ so that 
$\nu(l,3)$ is $2l+1$. The same computation in $D$ dimensions 
requires the computaiton of the degeneracy of $SO(D)$ representations
\begin{align}
\nu(l,D)&=\binom{l+D-1}{l}-\binom{l+D-3}{l-2} \ .
\label{degeneracio}
\end{align}

Given the explicit knowledge of all the
eigenvalues of  the reduced density matrix, we can 
also obtain a formula for the single copy entanglement Eq.(\ref{e1L}).
The largest eigenvalue of density matrix for two coupled harmonic 
oscillators is $(1-\xi)$. This largest eigenvalue of the density 
matrix $\rho_{out}$ will be the product of the largest eigenvalues of 
the density matrices which compound $\rho_{out}$,
\begin{align}
\rho^{(1)}_{out}&=\prod_{l\{m\}}\prod_{i=1}^{N-n}(1-\xi_{l\{m\},i}) \nonumber \\
&=\prod_{l=0}^{\infty}\left(\prod_{i=1}^{N-n}(1-\xi_{l\{m\},i})\right)^{\nu(l,D)} \ .
\end{align}
The single copy entanglement finally reads
\begin{equation}
E_1(\rho_L)=-\sum_{l=0}^{\infty}\nu(l,D) \left(\sum_{i=1}^{N-n} 
\log{(1-\xi_i)}\right)\ .
\label{single_copy2}
\end{equation}

\subsection{Perturbative computation for large angular momenta}
\label{Perturbative computation}
Note that our expressions for the entropy 
and the single copy entanglement depend on a final sum that ranges 
over all the values of angular momentum $l$. 
This sum may not be convergent as the radial discretization 
we have implemented is not a complete regularization of the field
theory. To be precise, 
the asymptotic dependence on $l$ should be under control in order
to correctly assess the convergence of the series. 

Let us note that, for $l\gg N$, the non diagonal elements of $K$ Eq.(\ref{Kdefinition}) 
are much smaller than the diagonal ones. These suggests the
possibility of setting up a  perturbative computation.

We split up the $K$ matrix in a diagonal $K_0$ and non diagonal $\lambda\eta$ 
matrices, where parameter $\lambda$ is just introduced to account for the order in
a perturbative expansion of the non-diagonal piece, 
\begin{equation}
K=K_0+\lambda\eta\ .
\end{equation}

This expansion is somewhat tedious and non illuminating. 
Technical details are presented in Appendix A.
The main observation is that the first contribution
$i=1$ out of every
set of $\xi_{l,\{m\},i}$ elements is relevant and it can
further be expanded as a series in $l^{-1}$, 
\begin{equation}
\xi \equiv \xi_{l,\{m\},1}=\frac{1}{l^4}\sum_{k=0}^5\frac{\xi_k}{l^k}+O(l^{-10})
\end{equation}   
We can then get the  entropy $S_{l\{m\}}$ .
\begin{align} 
S_{l\{m\}}&\simeq S_{l\{m\},1}= \sum_{k=1}^{\infty}  \left(\frac{1}{k}-\log (\xi )\right) \xi^k \nonumber \\
&=\frac{1}{l^4}\sum_{k=0}^5\frac{s_k+t_k\log{l}}{l^k}+O(l^{-10})\ ,
\end{align} 
where the coefficients $s_k$ and $t_k$ are defined in Appendix A.
A similar result for the single copy 
entanglement reads,
\begin{align}
E_1&\simeq\sum_{l=0}^{\infty}-\nu(l,D) \log{(1-\xi)}+O(l^{-10}) \nonumber \\
&=\sum_{l=0}^{\infty} \nu(l,D) \sum_{k=1}^{\infty} \frac{\xi^k}{k}\simeq\sum_{l=0}^{\infty} \nu \sum_{j=1}^{5}
 \frac{\kappa_j}{l^{4+j}}+O(l^{-10})\ ,
\label{single_copy2}
\end{align}
where $\kappa_j$ are the coefficients of the expansion given also in Appendix A.
Finally, using Eq.(\ref{degeneracio}) and defining $\tau_k\equiv\sum_{j=0}^k\nu_jt_{k-j}$ 
and $\sigma_k\equiv\sum_{j=0}^k\nu_j s_{k-j}$ 
where $\nu_j$ are the coefficients of the degeneracy expansion,
we determine the contribution to the total entropy, for $l=l_0\ldots \infty$, 
where $l_0$ is big enough such that the approximations are valid , 
\begin{align}
\label{entropytail}
  \Delta S &\simeq \sum_{j}^5\sigma_j\left( \zeta(6-D+j)-\sum_{l=1}^{l_0} 
  \frac{1}{l^{6-D+j}}\right) \nonumber \\
  &-\sum_{j}^5\tau_j\left(\zeta'(6-D+j)+\sum_{l=1}^{l_0} 
  \frac{\log l}{l^{6-D+j}}\right)
  &  
\end{align}
where $\zeta(n)$ is the Riemann Zeta function and $\zeta'(n)$ 
its derivative. 
Defining $\Lambda_k\equiv\sum_{j=0}^k\nu_j\kappa_{k-j}$, 
the single copy entanglement becomes,
\begin{align}
\Delta E_1&\simeq \sum_{j=0}^5\Lambda_j\left( \zeta(6-D+j)-\sum_{l=1}^{l_0} \frac{1}{l^{6-D+j}}\right)
\label{valor3}
\end{align} 

The above results show
that the sum over angular momenta $l$ converges for $D<5$.
A radial discretization of a scalar field theory
produces finite results for $D<5$ and needs further regularization
in orthogonal (angular) directions to the radius in higher dimensions. We will 
come back to this question later.

\subsection{Area law scaling}
The analysis of the scaling law obeyed by the
geometric entropy proceeds as follows. 
The analytical treatment of the chain of oscillators 
lead to the final sum over angular momenta in
Eq.(\ref{entropy2}). The computation of this
sum requires polynomial, rather than exponential, effort
as the size of the system increases. This justifies why
large systems are accessible within this approach. 
The eigenmodes $\xi_{l\{m\},i}$ are obtained by
diagonalization of matrices of order less than $N$. Finally the
tail of the sum over angular momenta is computed using the asymptotic
expressions given in Eq.(\ref{entropytail}).

\begin{figure}[hbt!]
\scalebox{0.7}{\input{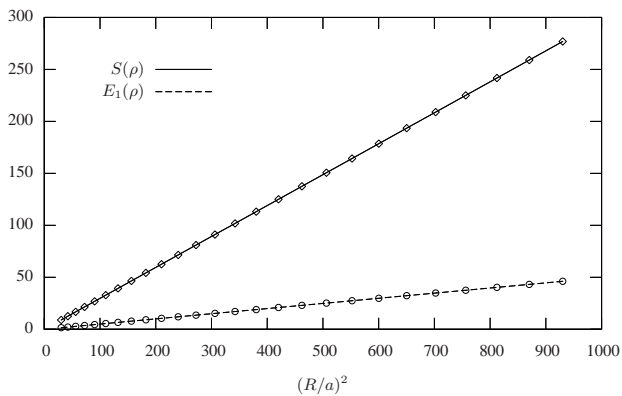}}
\caption{The entropy $S$ and the single copy entanglement $E_1$ 
resulting from tracing the ground state of a massless scalar field 
in three dimensions, over
the degrees of freedom inside a sphere of radius $R$.}
\label{figarealaw}
\end{figure}

We have computed the geometrical entropy and the single-copy
entanglement for different dimensionalities of the system.
Within the range $1<D<5$ we do observe the expected area law scaling
\begin{equation}
S=k_S(\mu,D,a,N) \left(\frac{R}{a}\right)^{D-1}\ ,
\end{equation}
as well as a similar scaling for the single copy entanglement
\begin{equation}
E_1=k_E(\mu,D,a,N)\left(\frac{R}{a}\right)^{D-1} \ ,
\end{equation}
where in all our considerations the lattice
spacing can be taken $a=1$. Fig.\ref{figarealaw} shows this  perfect
scaling for both measures of entanglement.

\begin{figure}[hbt!]
\scalebox{0.6}{\input{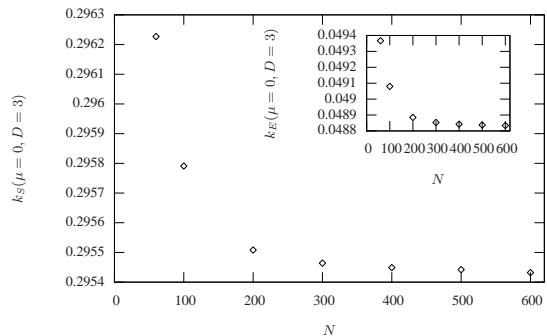}}
\caption{Coefficient for entropy the area law in $D=3$ as a function
of size of the system. Good stability is reached for $N=600$.
In the inset, the corresponding coefficient for the single copy 
area law is plotted.}
\label{arealawprefactor}
\end{figure}

The explicit pre-factor in the area law is regularization 
dependent but can be computed and compared with
previous analysis.
Fig.\ref{arealawprefactor} shows the result obtained for this pre-factor in 
the area law for the case of $D=3$ and 
$\mu=0$ as the size of the system increases.

Good stability is already reached for 
 $N=600$, 
where we recover the result of \cite{Sr93}
and complete it with the single copy entanglement
\begin{subequations}
\begin{align}
k_S(\mu=0,D=3, N \rightarrow \infty)&=0.295(1)\ , \\
k_E(\mu=0,D=3, N \rightarrow \infty)&=0.0488(1) \ .
\end{align}
\end{subequations}

Let us note that the ratio of the 
area law pre-factors for the entropy and the single
copy entanglement  is close to 6. This value is much larger
than the  factor of 2 computed to be the exact ratio
in one-dimensional critical systems
\cite{OLEC05}. We thus conclude that
the amount of entanglement that can be extracted from
a single copy of a system as compared to the asymptotic
value for infinite copies does decrease with the dimensionality.

\begin{figure}[hbt!]
 \scalebox{0.7}{\input{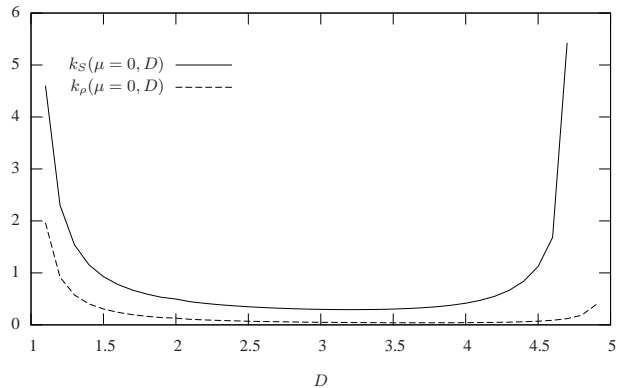}}
 \caption{Dependence of the geometric entropy and single copy
entanglement slopes, $k_S$ and $k_E$, 
on the dimension $D$ for a massless scalar field.
Note the divergence at $D=5$ due to the insufficient 
radial regularization
of the original field theory.}
\label{SEdimension}
\end{figure}

We can analyze in more detail the dependence of our two
measures of entanglement as a function of the 
dimensionality of the system. This is shown in Fig.\ref{SEdimension}.
for an $N=60$ and $5\le n\le30$ as a 
function of $R^{D-1}$ and we verify that the area law 
is observed for any value of the dimension $D$.

The robustness of area scaling law for arbitrary mass $\mu$
is also readily checked (see Fig.\ref{figmass}). The appearance
of a mass term in the Hamiltonian produces exponential
decays of correlators but does not affect the short-distance
entanglement which is ultimately responsible for the 
area law. This supports the idea that geometric entropy
comes from the local neighborhood of the surface separating
the region which is integrated out. The exponential decay
of massive modes is immaterial and their contribution
to the entanglement entropy is as
important as the one coming from massless modes.

Let us concentrate briefly in 
 the dependence of $k_S$ and $k_E$ on the dimension $D$. 
Those coefficients present divergences at $D=1$ and $D=5$ (see Fig.
\ref{SEdimension} ). 
The first one is due 
to the fact that in one dimension the strict power area law breaks down, 
since the limiting case carries
a logarithmic dependence. 
For $D\ge 5$, as we have shown before, the sum over partial 
waves does not converge. This is due to the fact that
we have regularized
the Hamiltonian using a radial 
lattice. This regularization is insufficient to handle
higher dimensional modes due to the increase of degrees of freedom
per radial shell. To avoid this problem, a more elaborated 
regularization of the initial  $D$-dimensional Hamiltonian is required.
Such a regularization will likely have to break the
rotational symmetry and will make the
computations rather involved.

\begin{figure}[hbt!]
 \scalebox{0.7}{\input{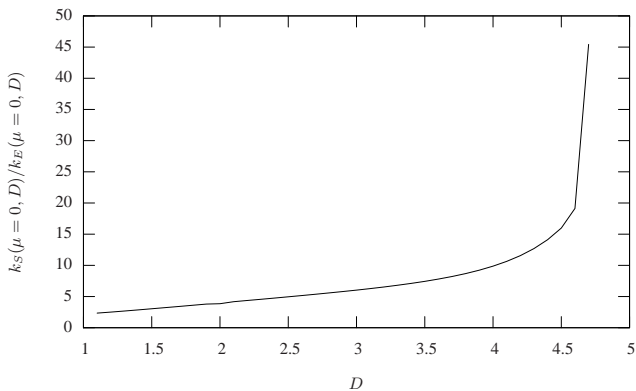}}
 \caption{Evolution of entropy to single-copy entanglement ratio $S/E_1$ as
a function of the dimension $D$. The line starts at a value of 2, as
demonstrated analytically in \cite{OLEC05} and grows monotonically. 
The higher the dimension is, the less entanglement is carried by
a single copy of the system as compared to many copies.}
\label{SoverEdimension}
\end{figure}

We observe in Fig.\ref{SoverEdimension} 
that the entropy to single-copy entanglement 
ratio verifies the expected limit 2, for $D$ tending to 1.

\subsection{Vacuum reordering}
Area law implies that entropy grows with the size of the system, that is, 
the eigenvalues of the density matrix, properly sorted
from the largest to the smallest, decay in a slower way for larger systems.
It has been  numerically shown in Ref.\cite{LLRV04}
that this order relation between systems of different length verifies the
strong condition of majorization, a fact proven analytically
for conformal field theories in Ref.\cite{Or05}. 
As the size of the system increases
from $L$ to $L'>L$, it is verified that $\rho_{L'}\prec \rho_L$,
where $\rho_L$ and $\rho_{L'}$ are the set of 
eigenvalues for the corresponding reduced density matrices.

Majorization relations characterize strong ordering. Every eigenvalue
changes in a way that is consistent with a set of
majorization constraints. We shall refer to this fact 
as {\sl vacuum reordering}. 

We show that the same underlying reordering of
the vacuum is present in any number of dimensions.
Unfortunately, a similar analytical treatment to the $D=1$ case 
is out of reach because the conformal group in $D>1$ 
is spanned by a finite number of generators.
As a consequence, there is no full control on the partition 
function of conformal field theory in $D>1$ dimensions, which 
could be used to generalize the one-dimensional theorem.

Vacuum reordering can be treated within our semi numerical approach.
From Eq.(\ref{tracedDM}) we see that 
the reduced density matrix of the exterior of a ball of radius $R$, can be expressed 
as a tensor product of simpler density matrices,
\begin{equation}
  \rho_{out}(R)=\prod_{l\{m\}} \rho_{l\{m\}}(R)=\prod_{l\{m\}} 
  \left( \prod_{i=1}^{N-n} \rho_{l\{m\},i}(R)\right) \ ,
  \label{densityM}
\end{equation}
where $\rho_{l\{m\}}$ is what we call $\rho_{out}$ 
in Sec.(\ref{chain of oscillators}) and $\rho_{l\{m\},i}$ are 
defined in the same section. 
A similar composition applies for  another size $R'$, 
\begin{equation}
  \rho_{out}(R')=\prod_{l\{m\}} \rho_{l\{m\}}(R')=\prod_{l\{m\}} 
  \left( \prod_{i=1}^{N-n'} \rho_{l\{m\},i}(R')\right) \ .
\end{equation}
It is shown as a lemma in Ref.\cite{LLRV04} that, if majorization relations 
are satisfied by each $\rho_{l\{m\}}(R)$ and $\rho_{l\{m\}}(R')$, 
they will be also satisfied by $\rho(R)$ and $\rho(R')$.
Note, though, that it is not possible to follow the same argument for $\rho_{l\{m\},i}(R)$ and 
$\rho_{l\{m\},i}(R')$ since $n \ne n'$.  To make dimensions agree,
we need to complete with identity
operators   
the smallest set. We then find that some majorization relations 
for the subparts are obeyed
in one sense, and the rest in the opposite one.
Thus, we construct the density matrices $\rho_{l\{m\}}(R)$ 
and $\rho_{l\{m\}}(R')$ doing the tensorial product of 
their components which are generated using 
Eq.(\ref{spectra}).
Once we have their eigenvalues we are ready to check that if $R < R'$, then
\begin{equation}
\rho_{out}(R') \prec \rho_{out}(R) \ ,
\end{equation}
which means by definition
\begin{equation}
 \sum_{i=1}^{k}p_i'\leq \sum_{i=1}^{k}p_i \, \, \, \, \forall \, k=1,\hdots,\infty \ 
\end{equation}
where $p_i$ and $p_i'$ are the eigenvalues of $\rho_{out}(R)$ 
and $\rho_{out}(R')$ respectively.
For the $l\sim N$ case, we have done a numerical computation with $N=60$ and 
truncating the vector of eigenvalues at the 50th element. 
Several dimensions $D$ and traced sizes $n$ have been studied, 
and all majorization relations are 
satisfied in all of them, as expected.
When $l\gg N$, we can use the analytical results of 
the Appendix A to check the same result.

\section{Entanglement loss along RG trajectories}

We shall now exploit the control achieved
on the eigenvalues of the reduced density matrix
in $D$ dimensions to study how entanglement evolves 
along renormalization group  transformations. 
This was studied for 
the quantum Ising model in Ref. \cite{LLRV04}. We shall
now add equivalent results for the set of harmonic oscillators
in $D$ dimensions. Results will turn out to be
qualitatively similar, reinforcing the concept
of entanglement loss along RG flows. 

The renormalization of a bosonic field is particularly
simple since the Hamiltonian only carries one coupling, 
namely the mass term. After a block transformation,
the rescaling of fields is used to make the kinetic
term to be normalized to $\frac{1}{2}$. 
The RG flow of the massive scalar field
reduces to an effective change of the mass.
That is, the study of the long distance behavior of
a correlator is viewed as taking a larger mass for the field,
modulo a scaling factor. This implies the existence of two fixed points  which are 
$\mu=0$ (ultraviolet, UV) and $\mu=\infty$ (infrared, IR). 
Since no other fixed point
is possible, the RG flow must be monotonic in $\mu$.

Entanglement loss comes along this flow. 
First, we study this change from a global perspective. 
We observe the obvious  global loss of entanglement. 
For $\mu=0$, geometric entropy grows with a slope 
$k_S(D,\mu=0)$ for the massless field
 and it is zero for the $\mu=\infty$ case. Thus,
\begin{equation}
S_{UV}\ge S_{IR} \ \ \ \ \forall R
\end{equation}
This result is related to the c-theorem as discussed
in Refs. \cite{Za86,CFL91,FL98,CH04, ZBFS}, which states
global irreversibility in the RG trajectory which interpolates 
between UV and IR fixed points.

\begin{figure}[hbt!]
\scalebox{0.7}{\input{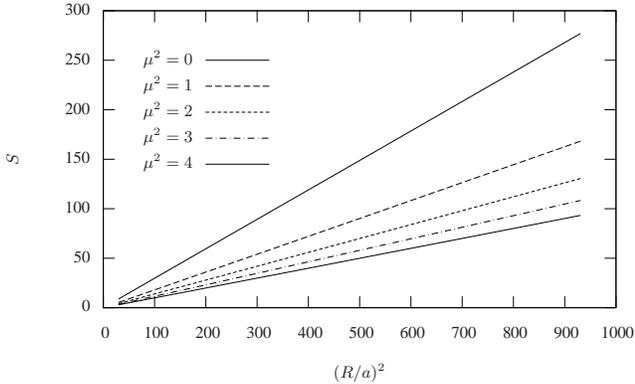}}
\caption{Geometric entropy $S$ for a sphere of radius $R$ 
in $D=3$ as a function of the  mass $\mu$. Note that larger masses
produce a smaller coefficient in the scaling are law. }
\label{figmass}
\end{figure}

On top of this global loss of entanglement, the geometric 
entropy obeys a  monotonic decrease along the RG flow. 
This behavior is illustrated for $D=3$ 
in Fig.\ref{figmass} the entropy 
for different masses where it is  seen that 
\begin{equation}
\mu' > \mu \Longrightarrow k_S(\mu') < k_S(\mu) .
\end{equation}
Thus, the system is 
more ordered as the mass increases. 

It is natural to pose the question if this order
relation verifies also stricter majorization relations,
that is, vacuum reordering.
Specifically, we analyze whether $\rho(\mu')$ and $\rho(\mu)$, 
the density matrices 
corresponding to the free bosonic model with masses $\mu' > \mu$ respectively,
obey $\vec{p}(\mu')$ and $\vec{p}(\mu)$.
\begin{equation}
\vec p(\mu) \prec \vec p(\mu') \ .
\end{equation}
\begin{figure}[hbt!]
\includegraphics{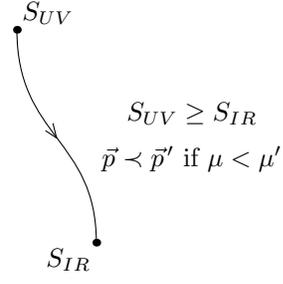}
\caption{Entanglement loss along the RG trajectories seen in
the space spanned by the eigenvalues of the reduced
density matrix.}
\label{eloss}
\end{figure}
Using similar arguments as in the previous section, 
we only need to check that each
$\rho_{l\{m\},i}(\mu)$ majorizes $\rho_{l\{m\},i}(\mu')$. Considering 
Eq.(\ref{spectra}), that means,
\begin{equation}
 \sum_{i=1}^{k}(1-\xi)\xi^i \leq \sum_{i=1}^{k}(1-\xi')\xi'^i \, \, \, \, \forall \, k=1,\hdots,\infty \ \ ,
\end{equation}
and therefore,
\begin{equation}
  (1-\xi^{k+1}) \leq (1-\xi'^{k+1}) \, \, 
\, \, \forall \, k=1,\hdots,\infty \ . 
\end{equation}
This happens if and only if $\xi' \leq \xi$. As in the previous section, 
we have verified this fact
numerically in the $l\sim N$ regime, and 
analytically using the perturbation calculus done in Appendix A.

It should be noted that monotonic loss of entanglement is
mandatory in such a simple model with a single 
parameter ($\mu$) controlling the flow. It is far from obvious
that such entropy loss is rooted in a such a subtle reordering
of the vacuum as the one dictated by majorization.

\section{Conclusions}

Area law scaling for the geometric entropy is present
in harmonic networks of arbitrary 
dimensions. This follows from a computation
that makes use of and analytical approach 
capable of making an analytical extension 
of the computation to arbitrary $D$, followed by a final
numerical resummation of angular momenta, whose
tail is
controlled analytically. 

A similar scaling law is observed for the single-copy entanglement.
This result suggests that entanglement, whatever measure
we use, scales with an area law due to the fact
that entanglement is concentrated on the surface
of the region which is traced out. The ratio of
single-copy entanglement to geometric entropy tends
to zero as the dimension of the network increases.

It is natural to interprete a change in the size of
the  subsystem which is traced out as well as
 any modification of
the parameters in the Hamiltonian as a probe on the
vacuum. Our explicit computations unveil ubiquous vacuum
reordering governed by majorization relations
of the vacuum state reduced density matrix eigenvalues.
Geometric entropy scaling is just one manifestation of 
this set of order relations.

The fact that finite PEPS support an area law scaling
makes them a natural tool to investigate regularized
quantum field theories.

\appendix 
\label{apendix}
\section{Perturbation theory}
We need to perform a perturbative computation for large momenta 
in order  to determine the contribution to 
the total entropy and single copy entanglement of all angular 
momentum modes.

We organize our computation in three parts. In the first 
part, we carry out  perturbation theory with matrices, following the 
same steps as explained in sec. (\ref{chain of oscillators}) when  
considering the aproximation $l\gg N$. This will produce
an analitical expression for the $\xi$'s parameters. The 
second part of the computation consists in Taylor expanding
the above results for $\xi$ in 
a series in $l^{-1}$. Finally, we will get the entropy and single
copy entanglement contributions, expanding the entropy 
and single copy modes in terms of $l^{-1}$ powers, 
and summing over $l$. In this sum 
$l$ take values from $l_0$ until infinity, where $l_0$ must be
suffitiently large, such that all aproximations done previously
are right.

\subsection{Computation of the $\xi$ parameter}
Let us recall that, for $l\gg N$, the non diagonal elements 
of $K$ in Eq. (\ref{Kdefinition}) 
are much smaller than the diagonal ones. That gives us the
possibility of setting up a  perturbative computation.

We split up the $K$ matrix in a diagonal $K_0$ and non diagonal $\lambda\eta$, 
matrices where $\lambda$ is just introduced to account for the order in
a perturbative expansion of the non-diagonal piece, 
\begin{equation}
   K=K_0+\lambda\eta \ .
\end{equation}
We will follow the steps described in Sec.\ref{chain of oscillators}. 
We expand $\Omega \equiv \sqrt{K}$ in its different contributions to order $\lambda$,
\begin{equation}
  \Omega = \Omega_0 + \lambda \epsilon + \lambda^2 \tilde{\epsilon} +
  \lambda^3 \hat{\epsilon}+ O(\lambda^4) \ .
\end{equation}
To get each term we impose the condition $\Omega^2=K$,
 \begin{align}
 (\Omega_0)_{ij} &= \Omega_i \delta_{ij} \nonumber \\
 (\epsilon)_{ij} &= \epsilon_i \delta_{i+1,j}+\epsilon_j \delta_{j+1,i}  \nonumber \\
 (\tilde{\epsilon})_{ij} &= \frac{\epsilon_i^2+\epsilon_{i-1}^2}{\Omega_i+\Omega_j}
  \delta_{ij}+ \frac{\epsilon_i \epsilon_{j-1}}{\Omega_i+\Omega_j} \delta_{i+2,j} \nonumber \\
  &+\frac{\epsilon_j \epsilon_{i-1}}{\Omega_i+\Omega_j} \delta_{i,j+2} \nonumber \\
 (\hat{\epsilon})_{ij}
 &=\frac{(\epsilon\tilde{\epsilon}+\tilde{\epsilon}\epsilon)_{ij}}{\Omega_i+\Omega_j} \ .
 \end{align}
where $\Omega_j$ and $\epsilon_j$ are defined since
 \label{Omega_definition}
 \begin{align}
 \Omega_j &\equiv \sqrt{\frac{l(l+D-2)}{j^2}+\omega_j} \ , \nonumber \\
 \omega_j &\equiv \left(1+\frac{1}{2j} \right)^{D-1}+  
 \left(1-\frac{1}{2j} \right)^{D-1}+\mu^2 \ ,  \nonumber \\
 \epsilon_j &\equiv  -\frac{j+\frac{1}{2}}{\sqrt{j(j+1)}}\frac{1}{\Omega_j+\Omega_{j+1}} \ .
 \end{align}
We structure $\Omega$ in three matrices $A$, $B$ and $C$,
 \begin{align}
 A &\equiv \Omega(1{\div}n,1{\div}n)=A_0+\lambda A_1 + \lambda^2 A_2 + O(\lambda^3) \nonumber\\
 B &\equiv \Omega(1{\div}n,n+1{\div}N)= \nonumber\\
   &=\lambda B_0+\lambda^2 B_1 + \lambda^3 B_2+ O(\lambda^4) \nonumber\\
 C &\equiv \Omega(n+1{\div}N,n+1{\div}N)=\nonumber \\
   &=C_0+\lambda C_1 + \lambda^2 C_2+O(\lambda^3) \ .
 \end{align} 
From these matrices, we define $\beta$ and $\gamma$ which we 
write in series of $\lambda$
\begin{align}
  \beta &\equiv \frac{1}{2} B^TA^{-1}B = \lambda^2\beta_0+\lambda^3\beta_1 
  +\lambda^4\beta_2+O(\lambda^4)\nonumber\\
  \gamma &\equiv C- \beta= \Omega_0 +  \lambda \epsilon + O(\lambda^2) \ ,
\end{align}
where,
 \begin{align}
\beta_0&=\frac{1}{2}B_0^T A^{-1}_0 B_0=\frac{\epsilon_n^2}{2\Omega_n} \delta_{i,1}\delta_{j,1}\nonumber\\
\beta_1&=\frac{1}{2}(B_1^T A^{-1}_0 B_0+B_0^T A^{-1}_0 B_1+B_0^T A^{-1}_1 B_0)  \nonumber \\
 &=-\frac{\epsilon_n^2}{2\Omega_n}\frac{\epsilon_{n+1}}{\Omega_{n}+\Omega_{n+2}}
 (\delta_{i,2}\delta_{j,1}+\delta_{i,1}\delta_{j,2}) \nonumber\\
 \beta_2&=\frac{1}{2}(B_2^T A^{-1}_0 B_0  +  B_0^T A^{-1}_0 B_2  + 
  B_0^T A^{-1}_2 B_0 \nonumber \\  
  &+  B_1^T A^{-1}_0 B_1  +  B_1^T A^{-1}_1 B_0  +  B_0^T A^{-1}_1 B_1)  \ .
 \end{align} 
We shall see later, that at 2n order perturbation in $\lambda$, 
only $(\beta_2)_{11}$ and $(\beta_2)_{22}$ of $\beta_2$ are necessary. Then,
 \begin{align}
(\beta_2)_{11}&=\frac{\epsilon_n^2}{2\Omega_n}\left\{\frac{\epsilon_{n-1}^2}{\Omega_n\Omega_{n-1}}+
\frac{\epsilon_{n-1}^2+\epsilon_{n}^2}{2\Omega_n^2}\right. \nonumber\\
&+\left.\frac{\Omega_{n}}{\Omega_{n-1}}\frac{\epsilon_{n-1}^2}{(\Omega_{n+1}+\Omega_{n-1})^2}
+\frac{2\epsilon_{n-1}^2}{\Omega_{n-1}(\Omega_{n+1}+\Omega_{n-1})}\right.\nonumber \\ 
&\left.+\frac{2}{\Omega_n(\Omega_n+\Omega_{n+1})}\left( \frac{\epsilon_{n+1}^2}{\Omega_n+\Omega_{n+2}}
\frac{\epsilon_{n+1}^2+\epsilon_{n}^2}{2\Omega_{n+1}}  \right. \right. \nonumber\\ 
&\left.\left.+  \frac{\epsilon_{n}^2+\epsilon_{n-1}^2}{2\Omega_{n}}+ 
\frac{\epsilon_{n-1}^2}{\Omega_{n+1}+\Omega_{n-1}}\right)\right\}  \nonumber \\
(\beta_2)_{22}&=\frac{\epsilon_n^2}{2\Omega_n}\frac{\epsilon_{n+1}^2}{(\Omega_n+\Omega_{n+2})^2}\ .
 \end{align} 
Let us diagonalize $\gamma$, 
\begin{equation}
\gamma_D = V\gamma V^T \ ,
\label{gamma_diag}
\end{equation}
where $V$ is an orthogonal matrix ($VV^T =\mathds{1}$).
Therefore, the eigenvalues are
\begin{align}
\det(\gamma-w\mathds{1}) &=\prod_{i=1}^{N-n}(\Omega_{n+i}-w)+O(\lambda^2)=0 \nonumber \\ 
\Rightarrow  w_i&=\Omega_{n+i}+O(\lambda^2) \ ,
\end{align}
and 
\begin{equation}
(\gamma_D)_{ij} =(\Omega_{n+i}+O(\lambda^2))\delta_{ij} \ .
\end{equation}
If we impose (\ref{gamma_diag}) over 
$V=V_0+\lambda V_1+\lambda^2 V_2+O(\lambda^3)$, we obtain,
 \begin{align}
 V_0&=\mathds{1} \nonumber \\
 (V_1)_{ij}&=\frac{\epsilon_{n+i}}{\Omega_{n+i}-\Omega_{n+j}} \delta_{i+1,j}+
 \frac{\epsilon_{n+j}}{\Omega_{n+j}-\Omega_{n+i}} \delta_{i,j+1} \nonumber \\
 (V_2)_{11}&=\frac{1}{2}\left( \frac{\epsilon_{n+1}}{\Omega_{n+1}-\Omega_{n+2}}\right)^2 \ .
 \end{align} 
Once we have $V$ and $\gamma_D$ we are able to compute 
$\beta'=\lambda^2(\beta'_0+\lambda \beta'_1+\lambda^2 \beta'_2+ O(\lambda^3) )$,
which is defined by,
\begin{equation}
\beta'\equiv \gamma_D^{-\frac{1}{2}}V \beta V^T\gamma_D^{-\frac{1}{2}} \ .
\end{equation}
Thus,
\begin{align}
\beta_0'&= (\gamma_D^{-\frac{1}{2}})_0 \beta_0(\gamma_D^{-\frac{1}{2}})_0 \nonumber \\
\beta_1'&= (\gamma_D^{-\frac{1}{2}})_0 \left[\beta_1+V_0\beta_1+\beta_1 V_0^T + V_1\beta_0+\beta_0 V_1^T 
 \right](\gamma_D^{-\frac{1}{2}})_0 \nonumber \\
\beta_2'&=(\gamma_D^{-\frac{1}{2}})_2 \beta_0 (\gamma_D^{-\frac{1}{2}})_0 
+(\gamma_D^{-\frac{1}{2}})_0 \beta_0 (\gamma_D^{-\frac{1}{2}})_2 \nonumber \\
&+(\gamma_D^{-\frac{1}{2}})_0 \left[
\beta_2+V_1\beta_1+\beta_1 V_1^T + V_2\beta_0+\beta_0 V_2^T \right.\nonumber \\
 &\left.+ V_1\beta_0 V_1^T \right](\gamma_D^{-\frac{1}{2}})_0
\end{align}
and therefore,
\begin{align}
 (\beta'_0)_{ij}&=\frac{\epsilon_n^2}{2\Omega_{n+1}\Omega_{n}}\delta_{i,1}\delta_{j,1} \nonumber\\
 (\beta'_1)_{ij}&=\frac{\epsilon_n^2}{2\Omega_{n+1}\Omega_{n}}
 \sqrt{\frac{\Omega_{n+1}}{\Omega_{n+2}}}
  \epsilon_{n+1}\nonumber \\
  &\times\left(\frac{1}{\Omega_{n+1}-\Omega_{n+2}}+\frac{1}{\Omega_{n}+\Omega_{n+2}}\right) 
  (\delta_{i,1}\delta_{j,2}+\delta_{i,2}\delta_{j,1}) \nonumber\\
 (\beta'_2)_{11}&= \frac{(\beta_2)_{11}}{\Omega_{n+1}}-\frac{\epsilon_n^2}{2\Omega_n\Omega_{n+1}} 
  \frac{\epsilon_{n+1}^2}{(\Omega_n-\Omega_{n-1})^2}   \nonumber  \\
  &- \frac{\epsilon_n^2}{2\Omega_n\Omega_{n+1}}
   \left(\frac{2\epsilon_{n+1}^2}{(\Omega_{n+2}+\Omega_{n})
        (\Omega_{n+1}-\Omega_{n+2})}\right. \nonumber \\
   &-\left.\frac{1}{2\Omega_{n+1}^2}\left(\epsilon_{n+1}^2
  \frac{\Omega_{n+1}+\Omega_{n+2}}{\Omega_{n+1}-\Omega_{n+2}}
  -\epsilon_{n}^2\frac{\Omega_{n}+\Omega_{n+1}}{\Omega_n}\right)\right)\nonumber\\
 (\beta'_2)_{22}&= \frac{\epsilon_{n}^2\epsilon_{n+1}^2}{2\Omega_{n+2}\Omega_{n}}
  \left(\frac{1}{\Omega_{n+1}-\Omega_{n+2}}+\frac{1}{\Omega_{n}+\Omega_{n+2}}\right)^2 \ . 
\end{align} 
It will be useful to write $\beta'$ in its matrix form,
\begin{equation}
\beta'=\lambda^2
\left( \begin{array}{cccc}
a_n +\lambda^2c_n& \lambda d_n & 0 & \ldots \\
\lambda d_n & \lambda^2 e_n & 0 & \ldots \\ 
0 & 0 & 0 & \ldots \\
\vdots & \vdots & \vdots & \ddots
\end{array} \right) +O(\lambda^5) \ ,
\end{equation}
where, 
\begin{align}
a_n&\equiv \frac{\epsilon_n^2}{2\Omega_{n+1}\Omega_{n}} \nonumber \\
d_n&\equiv a_n\sqrt{\frac{\Omega_{n+1}}{\Omega_{n+2}}}  \epsilon_{n+1}\left(\frac{1}{\Omega_{n+1}-\Omega_{n+2}}+\frac{1}{\Omega_{n}+\Omega_{n+2}}\right)
\end{align}
and $c_n$ and $e_n$ are respectively $(\beta'_2)_{11}$ and $(\beta'_2)_{22}$. 
We can observe now, that if we had not found the second order contribution of $(\beta')_{11}$ and $(\beta')_{22}$, we would not have been able to compute the eigenvalues of $\beta'$ to this order.\\
Diagonalizing $\beta'$, we find the eigenvalues $v_1$ and $v_2$,
\begin{align}
v_1&=\lambda^2\left(a_n+ \lambda^2\left(c_n+\frac{d^2}{a_n} \right) + O(\lambda^3) \right) \nonumber\\
v_2&=\lambda^4\left(e_n-\frac{d_n^2}{a_n} \right)+O(\lambda^5)=0+O(\lambda^5) \ ,
\end{align}
which allows us to compute the $\xi_i$'s parameters,
\begin{equation}
\xi_i=\frac{v_i}{1+\sqrt{1-v_i^2}} \ ,
\end{equation}
and which read
\begin{align}
\label{xi1}
\xi_1&=\frac{\lambda^2}{2}\left(a_n+ \lambda^2\left(c_n+\frac{d^2}{a_n} \right) + O(\lambda^3) \right) \nonumber\\
\xi_2&=0+O(\lambda^5)  \nonumber\\
\xi_i&=O(\lambda^7)\,\,\;\forall \,\,\; i>2  \ .
\end{align}
\subsection{Expansion of $\xi$ in terms of $l^{-1}$ powers}
We rename $\xi_1$ as $\xi$, and neglect the rest since at this order 
they are 0 and no contribute neither to the entropy nor to the single copy entanglement. 
We are interested in expanding $\xi$ in powers of $l^{-1}$. To do this, we have to expand first $\Omega_j$ and $\epsilon_j$, 
\begin{equation}
\Omega_n=l\sum_{i=0}^{9}\frac{\Omega_n^{(i)}}{l^i}+O(l^{-9}) \ ,
\end{equation}
where,
\begin{align}
\Omega_n^{(0)}&=\frac{1}{n} \nonumber \\
\Omega_n^{(1)}&=\frac{D-2}{2n} \nonumber \\
\Omega_n^{(2)}&=\frac{n\omega_n}{2}+\frac{(D-2)^2}{8n} \nonumber \\
\Omega_n^{(3)}&=\frac{(D-2)^3-4(D-2)n^2\omega_n}{16n} \nonumber \\
\Omega_n^{(4)}&=-\frac{5(D-2)^4-24(D-2)^2n^2\omega_n+16n^4\omega_n^2}{128n} \nonumber \\
\Omega_n^{(5)}&=\frac{7(D-2)^5}{256n}\nonumber \\
              &+\frac{-40(D-2)^2n^2\omega_n+48n^4\omega_n^2}{256n}  \\
&\vdots \qquad \ . \nonumber 
\end{align}
No more coefficients have been presented here since they have huge expressions 
and they don't shed any light on our arguments.
Using $\Omega_n$ we can obtain the expansion of $\epsilon_n$,
\begin{equation}
\epsilon_n=\frac{1}{l}\sum_{i=0}^{6}\frac{\epsilon_n^{(i)}}{l^i}+O(l^{-8}) \ ,
\end{equation}   
where,
\begin{align}
\epsilon_n^{(0)}&=\frac{(\eta)_{n,n+1}}{\Omega_n^{(0)}+\Omega_{n+1}^{(0)}} \nonumber \\
\epsilon_n^{(1)}&=-\epsilon^{(0)}_n\frac{\Omega^{(1)}_n+\Omega^{(1)}_{n+1}}
   {\Omega^{(0)}_n+\Omega^{(0)}_{n+1}} \nonumber \\
\epsilon_n^{(2)}&=\epsilon^{(0)}_n\left(\left(\frac{\Omega^{(1)}_n+\Omega^{(1)}_{n+1}}
{\Omega^{(0)}_n+\Omega^{(0)}_{n+1}}\right)^2-
  \frac{\Omega^{(2)}_n+\Omega^{(2)}_{n+1}}{\Omega^{(0)}_n+\Omega^{(0)}_{n+1}}\right)  \\
&\vdots \qquad \ . \nonumber 
\end{align}
Once we have $\Omega_n$ and $\epsilon_n$ in series of $l^{-1}$, we can expand $\xi$,
\begin{equation}
\xi=\frac{1}{l^4}\sum_{i=0}^{6}\frac{\xi_i}{l^i}+O(l^{-10}) \ ,
\end{equation}
with
\begin{align}
\xi_0&=\frac{(\epsilon^{(0)}_n)^2}{4\Omega^{(0)}_n+\Omega^{(0)}_{n+1}} \nonumber \\
\xi_1&=\epsilon^{(0)}_n\frac{2\epsilon^{(1)}_n\Omega^{(0)}_n\Omega^{(0)}_{n+1} - 
\epsilon^{(0)}_n\left(\Omega^{(1)}_n\Omega^{(0)}_{n+1} +\Omega^{(0)}_n\Omega^{(1)}_{n+1}\right)}
{4\left((\Omega^{(0)}_n)^2+(\Omega^{(0)}_{n+1})^2\right)}  \\
&\vdots \qquad \ . \nonumber 
\end{align}
Although $\xi$ depends on the number of oscillators which we trace out, we have omitted the subindex $n$ to simplify the notation. 

\subsection{The entropy}
The contribution to the entropy of a $(l,\{m\})$-mode becomes,
\begin{align}
S_{l\{m\}}&=-\log{(1-\xi_l)}-\frac{\xi_l}{1-\xi}\log{\xi_l} \nonumber \\
 &\simeq \sum_{k=1}^{\infty} \left(\frac{1}{k}-\log (\xi )\right) \xi^k \ .
\end{align}
If we substitute $\xi$,
\begin{equation}
S_{l\{m\}}=\frac{1}{l^4}\sum_{k=0}^5\frac{s_k+t_k\log{l}}{l^k}+O(l^{-10}) \ ,
\end{equation}
where,
\begin{align}
s_0&= \xi_0-\xi_0\log{\xi_0} \nonumber \\
s_1&= -\xi_1\log{\xi_0} \nonumber \\
s_2&= -\frac{\xi_1^2}{2\xi_0}-\xi_2\log{\xi_0} \nonumber \\  
s_3&= \frac{\xi_1^3-6\xi_0\xi_1\xi_2}{6\xi_0^2}-\xi_3\log{\xi_0} \nonumber \\
&\vdots \qquad \ , \nonumber 
\end{align}
and
\begin{align}
t_i&= 4\xi_i \,\,\,\,\,\, 0<i\leq3  \nonumber \\
t_4&= 4(\xi_0^2+\xi_4) \nonumber \\
t_5&= 4(2\xi_0\xi_1+\xi_5) \ .
\end{align}
To determine the contribution to the entropy of all modes with the
same $l$, we use the expansion of the degeneration, 
\begin{align}
\nu(l,D)&=\binom{l+D-1}{l}-\binom{l+D-3}{l-2} \nonumber \\
&=l^{D-2}\sum_{k=0}^{\infty}\frac{\nu_k(D)}{l^k} \ ,
\label{degeneracioA}
\end{align}
which allows us to sum over all the possible values of $\{m\}$,
\begin{equation}
 \sum_{\{m\}} S_{l\{m\}}= \nu(l,D)\, S_{l\{m\}} = 
 \sum_{i=0}^{5}\frac{\sigma_i\log l+\tau_i}{l^{6-D+i}}+O(l^{D-12})
\end{equation}
where $\tau_k\equiv\sum_{j=0}^k\nu_jt_{k-j}$ and $\sigma_k\equiv\sum_{j=0}^k\nu_js_{k-j}$. 
Finally, we can compute the contribution to total 
entropy, for $l$ from $l_0$ to $\infty$, where $l_0$ is 
big enough such that these aproximations are justified. 
\begin{align}
  \Delta S&\simeq \sum_{j}^5\sigma_j\left( \zeta(6-D+j)-\sum_{l=1}^{l_0} \frac{1}{l^{6-D+j}}\right)   \nonumber \\
  & - \sum_{j}^5\tau_j\left( \zeta'(6-D+j)+\sum_{l=1}^{l_0} 
  \frac{\log l}{l^{6-D+j}}\right) \ ,
\end{align}
being $\zeta(k)$ the Riemann Zeta function, and $\zeta'(k)$ its derivative.

\subsection{The single-copy entanglement}
We can do the same as for the entropy to find the contribution to the single-copy entanglement for large values of $l$.
First, we expand the contribution to the total single-copy entanglement of the ($l,\{m\}$)modes,
\begin{equation}
(E_1)_{l\{m\}}\simeq-\log{(1-\xi_l)}=\sum_{i=0}^{5}\frac{\kappa_i}{l^{4+i}}+O(l^{-10}) \ ,
\end{equation}
where, 
\begin{align}
\kappa_i&= \xi_i \,\,\,\,\,\, 0<i\leq3    \nonumber \\
\kappa_4&= (\frac{\xi_0^2}{2}+\xi_4) \nonumber \\
\kappa_5&= (\xi_0\xi_1+\xi_5)  \  .
\end{align}
Next, we sum for all possible values of $\{m\}$, using 
 Eq.(\ref{degeneracioA}),
\begin{align}
(E_1)_l=\nu(l,D)(E_1)_{l\{m\}}=\sum_{i=0}^{5}\frac{\Lambda_i}{l^{D-6+i}} \ ,
\end{align}
where $\Lambda_k\equiv\sum_{j=0}^k\nu_j\kappa_{k-j}$. 
Proceeding as before, we finally get
\begin{align}
E_1&\simeq \sum_{j=0}^5\Lambda_j\left( \zeta(6-D+j)-\sum_{l=1}^{l_0} \frac{1}{l^{6-D+j}}\right) \ .
\label{valor3}
\end{align} 

\emph{Acknowledgements.-} 
We would like to thank E. Rico who collaborated 
in the early discussion on majorization relations.
We would also like to acknowledge fruitful
discussions with M. Asorey, S. Iblisdir, R. Or\'us and J.M. Escartín.
Finally we are grateful to J. Eisert and M. Cramers. for a starting discussion on 
the ratio of $E_1/S$ in arbitrary dimensions.
This work has been supported by MEC (Spain), QAP (EU), Grup consolidat 
(Generalitat de Catalunya) and Universitat de Barcelona.

\end{document}